\documentclass[a4paper,12pt]{article}
\usepackage{amsmath,enumerate}
\usepackage{graphicx}

\begin{document}

\begin{center}

{\Large \bf $\mu\to e\gamma$ in a supersymmetric radiative neutrino
mass model}\\[20mm]

Raghavendra Srikanth Hundi\\
Department of Physics,\\
Indian Institute of Technology Hyderabad,\\
Kandi, Sangareddy - 502 285, Telangana, India.\\[5mm]
E-mail: rshundi@iith.ac.in \\[20mm]

\end{center}

\begin{abstract}
We have considered a supersymmetric version of the inert Higgs doublet model,
whose motivation is to explain smallness of neutrino masses and existence
of dark matter. In this supersymmetric model, due to the presence of discrete
symmetries, neutrinos acquire masses at loop level. After computing
these neutrino masses, in order to fit the neutrino oscillation data, we
have shown that by tuning some supersymmetry breaking soft parameters of
the model, neutrino
Yukawa couplings can be unsuppressed. In the above mentioned parameter
space, we have computed branching ratio of the decay $\mu\to e\gamma$. To
be consistent with the current experimental upper bound on $Br(\mu\to e\gamma)$,
we have obtained constraints on the right-handed neutrino mass of this
model.
\end{abstract}


\newpage
\section{Introduction}

There are many indications for physics beyond the standard model (SM)
\cite{bsm}. One among
them is the existence of non-zero neutrino masses \cite{neurev}.
Some of the indications
for new physics can be sucessfully explained in supersymmtric models
\cite{susy}.
For this reason, neutrino masses have been addressed in supersymmetry.
In a neutrino mass model, there is a possibility for
lepton flavor violation (LFV) \cite{lfvrev},
for which there is no direct evidence.
Experiments have put upper bounds on the branching ratios of these
LFV processes \cite{meg,lfvboun,pdg}.
Due to Glashow-Iliopoulos-Maiani cancellation mechanism, these processes
are highly suppressed in
the SM and the above mentioned upper bounds are obviously satisfied in it.
However, a signal for any LFV process with an appreciable branching ratio
gives a confirmation for new physics.

In this work, we study LFV processes of the form $\ell_i\to\ell_j\gamma$
in a supersymmetrized model for neutrino masses \cite{ma}. Here,
$\ell_i,i=1,2,3$, are charged leptons.
The above mentioned model arises after supersymmetrizing the
inert Higgs doublet model \cite{scoto,IH}. The inert Higgs doublet model
\cite{scoto} offers explanation for neutrino masses and dark matter.
In this model \cite{scoto}, dark matter is stable due to
an exact $Z_2$ symmetry
and the neutrinos acquire masses at 1-loop level.
This model has been extensively studied and some recent works on this
can be seen in \cite{IHrec}. Supersymmetrizing this model could bring new
features and it is done in \cite{ma}. In the supersymmetrization of the
inert Higgs doublet model \cite{ma}, the discreet symmetry is extended to
$Z_2\times Z_2^\prime$.
%In this model, apart from the usual Higgs superfields
%($\hat{H}_{u}$, $\hat{H}_u$), additional weak doublets ($\hat{\eta}_{1},
%\hat{\eta}_2$) and a
%singlet field ($\hat{\chi}$) are also proposed.
In this
model, dark matter can be multi-partite \cite{multi} due to the
presence of $R$-parity
and the $Z_2^\prime$ symmetry. Some variations of this model
are also presented in \cite{refs,st}.
In the model of \cite{ma}, gauge coupling unification
is possible by embedding it in a supersymmetric SU(5) structure \cite{su5}.
The origin of the discrete symmetry $Z_2\times Z_2^\prime$, which is
described above,
is also explained by realizing it as a residual symmetry from a U(1)
gauged symmetry \cite{u1ma}.

In this work we consider the model of \cite{ma} and present the expression
for neutrino masses, which arises from two 1-loop diagrams. We will
demonstrate that neutrino masses are tiny in this model if either the
neutrino Yukawa couplings are suppressed
or some certain soft parameters of the scalar potential are
fine-tuned. We consider the later case, in which the neutrino Yukawa couplings
can be ${\cal O}(1)$, and they can drive LFV processes such as
$\mu\to e\gamma$. In our work we assume flavor diagonal in the slepton
mass matricies as well as in the $A$-terms of sleptons. Hence, in our model,
lepton flavor violation is happening due to non-diagonal
Yukawa couplings. Under the above mentioned scenario, we compute
branching ratio for the decays $\ell_i\to\ell_j\gamma$.
Among these decays, we show that $\mu\to e\gamma$ can give stringent
constraints on model parameters, especially on right-handed neutrino mass.
Early calculations on $\mu\to e\gamma$ in a lepton number violating
supersymmetric model can be seen in \cite{ihl}.

In the model of \cite{ma}, apart from $\mu\to e\gamma$ there can also
be an LFV decay of $\mu\to 3e$. In a Type-II seesaw mechanism for neutrino
masses, the decay $\mu\to 3e$ can take place at tree level, due to the
presence of triplet Higgs boson. In our model
\cite{ma}, there are no triplet Higgses, hence the decay $\mu\to 3e$ will
take place at loop level. The current experimental upper limit on
$Br(\mu\to 3e)$ is $1\times 10^{-12}$ \cite{mu3e}, which is about two times
larger than that of $Br(\mu\to e\gamma)$. So we can expect
$Br(\mu\to e\gamma)$ to put
somewhat tighter constraints on model parameters than that due to
$Br(\mu\to 3e)$. Hence, in this work we focus on the computation of
$Br(\mu\to e\gamma)$. It may happen that $Br(\mu\to 3e)$ and
$Br(\mu\to e\gamma)$ may put some additional constraints on model parameters,
but we study these in a separate work.
%In Ref.[], neutrino masses and LFV processes are
%studied. However, we argue below that our work is different from that
%in Ref.[].
%The model of Ref.[st] assumes a theory at a high scale and reproduces the
%model of Ref.[ma] at a low scale. In Ref.[st], the $Z_2$ symmetry arises from
%the breakdown of an anomalous $U(1)_X$ symmetry. One difference between
%these two models is that the neutrino masses are determined by three
%and two 1-loop diagrams in Ref.[st] and Ref.[ma], respectively. Since
%we work in Ref. [ma], we show later that the smallness of neutrino masses
%is governed by the smallness of some soft parameters in the scalar potential
%of this model. Whereas, in Ref.[], neutrino masses are small because of
%small trilinear couplings of $\hat{H}_d\hat{\eta}_2\hat{\chi}$ and
%$\hat{H}_u\hat{\eta}_1\hat{\chi}$. Moreover, we will show later
%that the expression for the LFV process $\mu\to e\gamma$ in our work
%is different from that of Ref.[st]. This difference might have arised
%since the model Ref.[st] has different origin compared to Ref.[ma].

This paper is organized as follows. In the next section, we describe the
model of \cite{ma}. In section 3, we present the expressions for neutrino
masses and branching ratios for the decays $\ell_i\to\ell_j\gamma$.
In section 4, we give
neumerical results on neutrino masses and $\mu\to e\gamma$. We conclude in
section 5.

\section{The model}

The model of Ref.\cite{ma} is an extension of minimal supersymmetric standard
model (MSSM).
The additional superfields of this model are as follows:
(i) three right-handed neutrino fields, $\hat{N}_i$, $i=1,2,3$, (ii)
two electroweak doublets $\hat{\eta}_1=(\hat{\eta}_1^0,\hat{\eta}_1^-)$,
$\hat{\eta}_2=(\hat{\eta}_2^+,\hat{\eta}_2^0)$, (iii)
a singlet field $\hat{\chi}$. Under the electroweak gauge group
SU(2)$_L\times$U(1)$_Y$, the charges of these additional superfields are
given in Table 1.
\begin{table}[!h]
\begin{center}
\begin{tabular}{c|cccc}\hline\hline
Field & $\hat{N}_i$ & $\hat{\eta}_1$ & $\hat{\eta}_2$ & $\hat{\chi}$ \\\hline
%& & $(\hat{\eta}_1^0,\hat{\eta}_1^-)$ & $(\hat{\eta}_2^+,\hat{\eta}_2^0)$ &
%\\\hline
SU(2)$_L\times$U(1)$_Y$ & (1,0) & (2,-1/2) & (2,1/2) & (1,0) \\\hline\hline
\end{tabular}
\end{center}
\caption{Charge assignments of additional superfields of the model
under the electroweak gauge group.}
\end{table}
The model of Ref.\cite{ma} contains discrete symmetry $Z_2\times Z_2^\prime$,
under which all the quark and Higgs superfields can be taken to be even.
The leptons and the additional fields described above are charged
non-trivially under this discrete symmetry \cite{ma}. The purpouse
of this symmetry is to disallow the Yukawa term $\hat{L}_i\hat{H}_u\hat{N}_j$
in the superpotential of the model, and as a result the neutrino
remains massless at tree level. Here,
$\hat{L}_i=(\hat{\nu}_i,\hat{\ell}_i)$, $i=1,2,3$, are
the lepton doublet superfields. The singlet charged lepton superfield
is represented by $\hat{E}^c_i$, $i=1,2,3$.
We denote up- and down-type Higgs superfields
as $\hat{H}_u$ and $\hat{H}_d$ respectively.

The superpotential of our model consisting of electroweak fields can be
written as \cite{ma}
\begin{eqnarray}
W&=& (Y_E)_{ij}\hat{L}_i\hat{H}_d\hat{E}^c_j+
(Y_\nu)_{ij}\hat{L}_i\hat{\eta}_2\hat{N}_j+\lambda_1\hat{H}_d\hat{\eta}_2
\hat{\chi}+\lambda_2\hat{H}_u\hat{\eta}_1\hat{\chi}+
\nonumber \\
&&\mu \hat{H}_u\hat{H}_d+\mu_\eta\hat{\eta}_2
\hat{\eta}_1+\frac{1}{2}\mu_\chi\hat{\chi}\hat{\chi}+\frac{1}{2}M_{ij}
\hat{N}_i\hat{N}_j
\label{eq:W}
\end{eqnarray}
Here, there is a summation over indices $i,j$ which run from 1 to 3.
The first and second terms in the above equation are Yukawa terms
for charged leptons and neutrinos, respectively. But,
as described before, $\hat{\eta}_2$ is odd under the discrete symmetry
of the model and hence the scalar component of it does not acquire vacuum
expectation value (vev) \cite{ma}. So neutrinos are still massless at
tree level.
Apart from the superpotential of Eq. (\ref{eq:W}), we should consider
the scalar potential. The relavant terms in the scalar potential
are given below.
\begin{eqnarray}
V &=& (m^2_L)_{ij}\tilde{L}_i^\dagger\tilde{L}_j+m^2_{\eta_1}\eta_1^\dagger
\eta_1+m^2_{\eta_2}\eta_2^\dagger\eta_2+m^2_\chi\chi^*\chi+
(m^2_N)_{ij}\tilde{N}^*_i\tilde{N}_j+
\nonumber \\
&& \left[(AY_\nu)_{ij}\tilde{L}_i\eta_2\tilde{N}_j+(A\lambda)_1H_d\eta_2\chi
+(A\lambda)_2H_u\eta_1\chi \right.
\nonumber \\
&&\left. +b_\eta\eta_2\eta_1+\frac{1}{2}b_\chi\chi\chi
+\frac{1}{2}(b_M)_{ij}\tilde{N}_i\tilde{N}_j+{\rm c.c.}\right].
\label{E:spot}
\end{eqnarray}

As we have explained before that our motivation is to study LFV processes
in the above described model. The LFV processes can be driven by charged
sleptons. For instance, the off-diagonal elements of soft parameters,
$(m^2_L)_{ij}$, can drive LFV processes. Similarly, we can write soft mass
terms for singlet charged sleptons, $\tilde{E}_i,i=1,2,3$, in the
scalar potential. Also, there can exist
$A$-terms connecting $\tilde{L}_i$ and $\tilde{E}_j$. The off-diagonal
terms of the above mentioned soft terms can drive LFV processes, which actually
exist in MSSM. Since our model \cite{ma} is an extension of MSSM, we
are interested in LFV processes generated by the additional fields
of this model. Hence, we assume that the off-diagonal terms of the
soft terms, which are described above, are zero.

For simplicity, we assume that the parameters of the superpotential and
scalar potential of our model are real. Then, by an orthogonal transformation
among the neutrino superfields, $\hat{N}_i$, we can make the
the following parameters to be diagonal, which are given below.
\begin{equation}
M_{ij}=M_i\delta_{ij},\quad (m_N^2)_{ij}=(m_N^2)_i\delta_{ij},\quad
(b_M)_{ij}=(b_M)_i\delta_{ij}
\end{equation}
By going to an appropriate basis of $\hat{L}_i$ and $\hat{E}_j$, we can
get the Yukawa couplings for charged leptons to be diagonal.
After doing this, we are left with no freedom and hence the neutrino
Yukawa couplings, $(Y_\nu)_{ij}$, can be non-diagonal. These
non-diagonal Yukawa couplings can drive LFV processes such as
$\ell_i\to\ell_j\gamma$.
These LFV processes are driven at the 1-loop level, which we describe
in the next section. As explained before, neutrinos also acquire masses
at 1-loop level in this model \cite{ma}. To calculate these loop diagrams
we need to know the mass eigenstates of the scalar and fermionic partners
of the fields shown in Table 1, since these fields enter into the
loop processes. Expressions for these mass eigenstates are given in
Ref.\cite{akot}. However, our notations and conventions are different from
that of Ref.\cite{akot}. Hence, for the sake of completeness we present
them below.

The charged components of $\hat{\eta}_1,\hat{\eta}_2$ can be
fermionic and scalar, which can be written as
$(\tilde{\eta}_1^-,\tilde{\eta}_2^+)$ and $(\eta_1^-,\eta_2^+)$,
respectively. The two charged fermions
represent chargino-type fields, whose mass is $\mu_\eta$.  Whereas,
the charged scalars, in the basis $\Phi_+^{\rm T}=(\eta_2^+,\eta_1^{-*})$,
will have a mass matrix which is given below.
\begin{equation}
{\cal L}\ni -\Phi_+^\dagger
\left(\begin{array}{cc}
\mu_\eta^2+m_{\eta_2}^2+\frac{g^2-g^{\prime 2}}{4}v^2\cos(2\beta) & b_\eta \\
b_\eta & \mu_\eta^2+m_{\eta_1}^2-\frac{g^2-g^{\prime 2}}{4}v^2\cos(2\beta)
\end{array}\right)\Phi_+
\end{equation}
Here, $g,g^\prime$ are the gauge couplings of SU(2)$_L$ and U(1)$_Y$,
respectively. $\beta$ is defined as $\tan\beta =\frac{v_2}{v_1}=
\frac{\langle H_u^0\rangle}{\langle H_d^0\rangle}$ and $v^2=v_1^2+v_2^2$.
We can diagonalize the above mass matrix by taking $\Phi_+$ as
\begin{eqnarray}
\Phi_+=\left(\begin{array}{cc}
\cos\theta & -\sin\theta \\
\sin\theta & \cos\theta
\end{array}\right)
\left(\begin{array}{c}
\eta^+_{m2}\\\eta^+_{m1}\end{array}\right),
\quad
\tan 2\theta = \frac{2b_\eta}{m^2_{\eta_2}-m^2_{\eta_1}+(g^2-g^{\prime 2})
v^2\cos(2\beta)/2}
\end{eqnarray}
Here, $\eta^+_{m1}$ and $\eta^+_{m2}$ are mass eigenstates of the
charged scalar fields and we denote their mass eigenvalues by
$m_{1+}$ and $m_{2+}$, respectively.

The neutral fermionic and scalar components of
$\hat{\eta}_1,\hat{\eta}_2,\hat{\chi}$ can be written as
$\Psi^{\rm T}=(\tilde{\eta}_1^0,\tilde{\eta}_2^0,\tilde{\chi})$
and $\Phi_0^{\rm T}=(\eta_1^0,\eta_2^0,\chi)$, respectively.
The neutral fermionic fields will have a mixing mass matrix,
which is given below.
\begin{equation}
{\cal L} \ni -\frac{1}{2}\Psi^{\rm T}M_\eta\Psi,
\quad
M_\eta = \left(\begin{array}{ccc}
0 & -\mu_\eta & -\lambda_2v_2 \\
-\mu_\eta & 0 & \lambda_1v_1 \\
-\lambda_2v_2 & \lambda_1v_1 & \mu_\chi \end{array}\right)
\end{equation}
The above mixing matrix can be diagonalized by an orthogonal matrix as
\begin{equation}
U_\eta^{\rm T}M_\eta U_\eta = {\rm diag}(m_{\tilde{\eta}_1},m_{\tilde{\eta}_2},
m_{\tilde{\eta}_3})
\end{equation}

The neutral scalar fields of $\Phi_0$ can be written as
\begin{equation}
\Phi_0=\frac{1}{\sqrt{2}}\Phi_R+\frac{i}{\sqrt{2}}\Phi_I=
\frac{1}{\sqrt{2}}\left(\begin{array}{c}
\eta^0_{1R}\\\eta^0_{2R}\\\chi_R
\end{array}\right)+\frac{i}{\sqrt{2}}
\left(\begin{array}{c}
\eta^0_{1I}\\\eta^0_{2I}\\\chi_I
\end{array}\right)
\end{equation}
The mixing matrix among these fields can be written as
\begin{equation}
{\cal L}\ni -\frac{1}{2}\Phi_R^{\rm T}m_{\eta_R}^2\Phi_R
-\frac{1}{2}\Phi_I^{\rm T}m_{\eta_I}^2\Phi_I
\end{equation}
Here, the mixing matrices $m_{\eta_R}^2,m_{\eta_I}^2$ can be obtained from the
following matrix
\begin{eqnarray}
m_\eta^2(\epsilon)&=&\left(\begin{array}{ccc}
m_{11}^2 & m^2_{12} & m^2_{13} \\
m^2_{12} & m^2_{22} & m^2_{23} \\
m^2_{13} & m^2_{23} & m^2_{33}
\end{array}\right),
\quad
m_{11}^2= \mu_\eta^2+m_{\eta_1}^2+\lambda_2^2v_2^2+\frac{g^2+g^{\prime 2}}
{4}v^2\cos(2\beta),
\nonumber \\
m_{22}^2&=& \mu_\eta^2+m_{\eta_2}^2+\lambda_1^2v_1^2-\frac{g^2+g^{\prime 2}}
{4}v^2\cos(2\beta),
\quad
m_{33}^2= \mu_\chi^2+m_\chi^2+\lambda_1^2v_1^2+\lambda_2^2v_2^2+\epsilon
b_\chi ,
\nonumber \\
m_{12}^2&=& -\lambda_1\lambda_2v_1v_2-\epsilon b_\eta ,
\quad
m_{13}^2= -\lambda_1v_1\mu_\eta -\lambda_2v_2\mu_\chi -\epsilon
[(A\lambda)_2v_2-\mu \lambda_2v_1]
\nonumber \\
m_{23}^2&=& \lambda_1v_1\mu_\chi +\lambda_2v_2\mu_\eta +\epsilon
[(A\lambda)_1v_1-\mu \lambda_1v_2]
\label{Eq:m2eta}
\end{eqnarray}
Here, $\epsilon$ can take $+1$ or $-1$. We have $m_{\eta_R}^2=m_\eta^2(+1)$ and
$m_{\eta_I}^2=m_\eta^2(-1)$. These two mixing mass matrices can be
diagonalized by
orthogonal matrices $U_R$ and $U_I$, which are defined below.
\begin{equation}
U_R^{\rm T}m_{\eta_R}^2U_R={\rm diag}(m^2_{\eta_{R1}},m^2_{\eta_{R2}},
m^2_{\eta_{R3}}),
\quad
U_I^{\rm T}m_{\eta_I}^2U_I={\rm diag}(m^2_{\eta_{I1}},m^2_{\eta_{I2}},
m^2_{\eta_{I3}})
\end{equation}

At last, the fermionic and scalar components of right-handed neutrino
superfields, $\hat{N}_i$, can be donted by $N_i$ and $\tilde{N}_i$,
respectively. The fermionic components have masses $M_i$. The scalar
components can be decomposed into mass eigenstates as
\begin{equation}
\tilde{N}_i=\frac{1}{\sqrt{2}}\left(\tilde{N}_{Ri}+i\tilde{N}_{Ii}\right)
\label{Eq:sneu}
\end{equation}
The mass-squares of $\tilde{N}_{Ri}$ and $\tilde{N}_{Ii}$, respectively, are
given below.
\begin{equation}
m_{Ri}^2=M_i^2+(m_{N}^2)_i+(b_{M})_i,\quad
m_{Ii}^2=M_i^2+(m_{N}^2)_i-(b_{M})_i
\label{Eq:m2RI}
\end{equation}

\section{Neutrino masses and LFV processes}

As described before that in the model of Ref.\cite{ma} neutrinos are
massless at tree level due to the presence of the discrete symmetry
$Z_2\times Z_2^\prime$. However, in this model
neutrinos acquire masses at 1-loop level, whose diagrams are shown
in Figure 1 \cite{ma}.
\begin{figure}[!h]
\begin{center}

\includegraphics[width=6in]{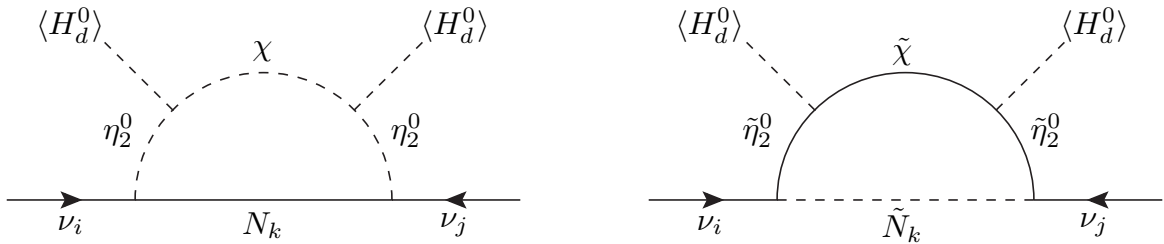}

\end{center}
\caption{Radiative masses for neutrinos.}
\end{figure}
After computing these 1-loop diagrams, we have found the following
mass matrix for neutrinos.
\begin{eqnarray}
\left(m_\nu\right)_{ij}&=&\sum_{k,l=1}^3\frac{(Y_\nu)_{ik}(Y_\nu)_{jk}}
{16\pi^2}M_k
\left[[U_R(2,l)]^2\frac{m_{\eta_{Rl}}^2}{m_{\eta_{Rl}}^2-M_k^2}\ln
\frac{m_{\eta_{Rl}}^2}{M_k^2}-
[U_I(2,l)]^2\frac{m_{\eta_{Il}}^2}{m_{\eta_{Il}}^2-M_k^2}\ln
\frac{m_{\eta_{Il}}^2}{M_k^2}\right]
\nonumber \\
&&+\sum_{k,l=1}^3\frac{(Y_\nu)_{ik}(Y_\nu)_{jk}}{16\pi^2}[U_\eta(2,l)]^2
m_{\tilde{\eta}_l}
\left[\frac{m_{Rk}^2}{m_{Rk}^2-m_{\tilde{\eta}_l}^2}\ln\frac{m_{Rk}^2}
{m_{\tilde{\eta}_l}^2}-
\frac{m_{Ik}^2}{m_{Ik}^2-m_{\tilde{\eta}_l}^2}\ln\frac{m_{Ik}^2}
{m_{\tilde{\eta}_l}^2}\right]
\label{Eq:mnu}
\end{eqnarray}
It is to be noticed that the first and second lines of the above
equation arises from the left- and right-handed diagrams of Figure 1.

In our work we assume supersymmetry breaking to be around 1 TeV. Hence,
we can take all the supersymmetric (SUSY) particle masses to be around few
hundred GeV. With this assumption, we can estimate the
neutrino Yukawa couplings by requiring that the neutrino mass scale to be
around 0.1 eV \cite{neurev}. With this requirement, we have found that
$(Y_\nu)_{ij}\sim 10^{-5}$.
Here there are six different Yukawa couplings, which need to be suppressed
to ${\cal O}(10^{-5})$. This could be one possibility in this model
in order to explain the
correct magnitude for neutrino masses. However, in this case, since
the Yukawa couplings are suppressed, LFV processes such as
$\ell_i\to\ell_j\gamma$ would also be suppressed. These LFV processes
will be searched in future experiments \cite{fulfv}, hence
it is worth to consider the case where these processes can have substancial
contribution in this model. In otherwords, we have to look for a parameter
region where we can have $(Y_\nu)_{ij}\sim{\cal O}(1)$.

From Eq. (\ref{Eq:mnu}), it can observed that each diagram of Figure 1
contribute positive and negitive quantities to the neutrino mass
matrix. Without suppressing Yukawa couplings, by fine-tuning
the masses of SUSY particles, we may achieve
partial cancellation between the positive and negative contributions
of Eq. (\ref{Eq:mnu}) and endup with tiny masses for neutrinos. To
demonstrate this explicitly, using Eq. (\ref{Eq:m2RI}), we can notice
that in the limit $(b_M)_i\to 0$ we get $m_{Ri}^2- m_{Ii}^2\to 0$,
and hence the second line of Eq. (\ref{Eq:mnu}) would give tiny contribution.
The first line of Eq. (\ref{Eq:mnu}) can give very small value in the
following limiting process: $U_R(2,l)-U_I(2,l)\to 0$ and
$m_{\eta_{Rl}}-m_{\eta_{Il}}\to 0$. To achieve this limiting process
we have to make sure that the elements of the matrices $m_{\eta_R}^2$ and
$m_{\eta_I}^2$ are close to each other. From the discussion around Eq.
(\ref{Eq:m2eta}), we can observe that the elements of $m_{\eta_R}^2$ and
$m_{\eta_I}^2$ can differ by quantities which are proportional to $\epsilon$.
These quantities depend on the following parameters:
$b_\chi$, $b_\eta$, $(A\lambda)_1$ and $(A\lambda)_2$. By taking the
following limit:
$(A\lambda)_{1}-\lambda_{1}\mu v_2/v_1\to 0$,
$(A\lambda)_{2}-\lambda_{2}\mu v_1/v_2\to 0$, $b_\eta\to 0$, $b_\chi\to 0$,
we can get tiny contribution from the first line of Eq. (\ref{Eq:mnu}).
To sum up the above discussion, without suppressing the neutrino Yukawa
couplings we can fine-tune the below seven paramters, in order to
get very small neutrino masses in this model.
\begin{equation}
(b_M)_i,i=1,2,3,\quad b_\eta,\quad b_\chi,\quad (A\lambda)_1,\quad
(A\lambda)_2
\label{Eq:para}
\end{equation}
Apparently, the above parameters are SUSY breaking soft parameters
of the scalar potential of this model. A study of neutrino masses depending
on SUSY breaking soft parameters can be seen in \cite{af}.

In the previous paragraph we have argued that Majorana masses for neutrinos
are vanishingly small when we fine tune certain soft parameters of the model.
We can understand these features from symmetry arguments.
For instance, when lepton number
is conserved, neutrinos cannot have Majorana masses. For lepton number,
we can propose a group U(1)$_L$, under which the following fields are assigned
the corresponding charges and the rest of the superfields are singlets.
\begin{equation}
\hat{L}_i\mapsto +1,\quad\hat{E}^c_i\mapsto -1,\quad\hat{N}_i\mapsto -1
\end{equation}
With the above mentioned charges, we can see that the last term in Eqs.
(\ref{eq:W}) and (\ref{E:spot}) are forbidden. In fact, in the limit
$M_i\to 0$ and $(b_M)_i\to 0$, the two diagrams of Figure 1 give zero masses
to neutrinos. Hence, in order to get Majorana masses for neutrinos, we have
softly broken the lepton number symmetry. Now, even if we have $M_i\ne 0$,
we have described in the previous paragraph that the left-handed diagram of
Figure 1 can still give vanishingly small masses by fine tuning some
soft parameters.
This suggests that apart from U(1)$_L$ there can exist some additional
symmetries. Suppose we set $(A\lambda)_{1}v_1-\lambda_{1}\mu v_2=0$,
$(A\lambda)_{2}v_2-\lambda_{2}\mu v_1=0$. Then, as argued previously that
the left-handed diagram of Figure 1 gives zero neutrino masses for $b_\eta\to 0$
and $b_\chi\to 0$, even if $M_i\ne 0$. This case can be understood by
proposing additional symmetry U(1)$_\eta$, under which the following fields
have non-trivial charges and the rest of the fields are singlets.
\begin{equation}
\hat{L}_i\mapsto +1,\quad\hat{E}^c_i\mapsto -1,\quad\hat{\eta}_1\mapsto -1,
\quad\hat{\eta}_2\mapsto -1,\quad\hat{\chi}\mapsto +1
\end{equation}
Using the above charges, we can notice that $\mu_\eta$-, $\mu_\chi$-terms
in Eq. (\ref{eq:W}) and $b_\eta$-, $b_\chi$-terms in Eq. (\ref{E:spot}) are
forbidden. Thus, the additional symmetry U(1)$_\eta$ can forbid the Majorana
masses for neutrinos in the left-handed diagram of Figure 1. Finally, one may
ask how the relations $(A\lambda)_{1}v_1-\lambda_{1}\mu v_2=0$,
$(A\lambda)_{2}v_2-\lambda_{2}\mu v_1=0$ can be satisfied. In these two
relations, SUSY breaking soft masses are related to SUSY conserving mass
$\mu$. These relations may be achieved my proposing certain symmetries
in the mechansim
for SUSY breaking, which is beyond the reach of our present work.

Previously, we have motivated a parameter region where
the neutrino Yukawa couplings can be ${\cal O}(1)$. For these values
of neutrino Yukawa couplings, LFV processes such as $\ell_i\to\ell_j\gamma$
can have substantical contribution in our model, and worth to compute them.
The Feynman diagrams for $\ell_i\to\ell_j\gamma$ are given in Figure 2.
\begin{figure}[!h]
\begin{center}

\includegraphics[width=6in]{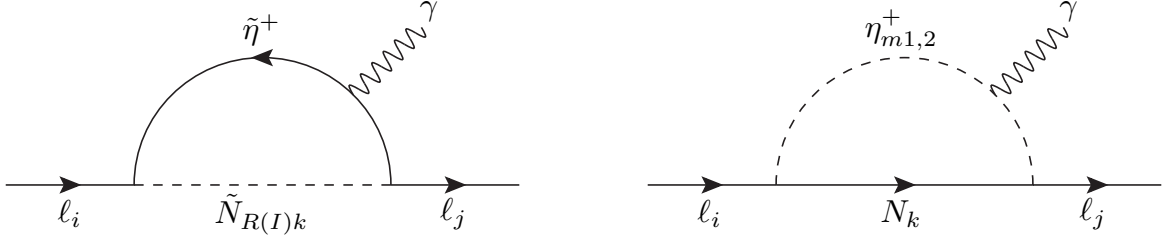}

\end{center}
\caption{Lepton flavor violating decays of the form $\ell_i\to\ell_j\gamma$.}
\end{figure}

The general form of the amplitude for $\ell_i\to\ell_j\gamma$ is as follows.
\begin{equation}
{\cal M}=e\epsilon^*_\mu(q)\bar{u}_j(p-q)\left[A_L^{(ij)}\frac{1-\gamma_5}{2}
+A_R^{(ij)}\frac{1+\gamma_5}{2}\right]i\sigma^{\mu\nu}q_\nu u_i(p)
\label{Eq:amp}
\end{equation}
It is to be noted that in the above equation, there is no summation over
the indices $i,j$. The quantities $A_{L,R}^{(ij)}$ of the above
equation can be found from the 1-loop diagrams of Figure 2, which we
have given below.
\begin{eqnarray}
A_L^{(ij)}&=&A^{(ij)}m_j,\quad A_R^{(ij)}=A^{(ij)}m_i,
\nonumber \\
A^{(ij)}&=&\sum_{k=1}^3\frac{(Y_\nu)_{ik}(Y_\nu)_{jk}}{16\pi^2}
\left\{\frac{1}{4\mu_\eta^2}\left[f_2(x_{Rk})+f_2(x_{Ik})\right]
-\left[\cos^2\theta\frac{f_2(x_{k2})}{2m^2_{2+}}+\sin^2\theta\frac{f_2(x_{k1})}
{2m^2_{1+}}\right]\right\},
\nonumber \\
x_{Rk}&=&\frac{m_{Rk}^2}{\mu_\eta^2},\quad x_{Ik}=\frac{m_{Ik}^2}{\mu_\eta^2},
\quad x_{k2}=\frac{M_k^2}{m^2_{2+}},\quad x_{k1}=\frac{M_k^2}{m^2_{1+}},
\nonumber \\
f_2(x)&=&\frac{1}{(1-x)^4}\left[\frac{1}{6}-x+\frac{1}{2}x^2+\frac{1}{3}
x^3-x^2\ln(x)\right].
\end{eqnarray}
From the above expressions, we can notice that in the curly brackets of
$A^{(ij)}$, the first two and the last two terms are arising from the
left- and right-handed diagrams of Figure 2, respectively. Moreover,
there is a relative minus sign in the contribution from these two diagrams.

Among the various decays of the form $\ell_i\to\ell_j\gamma$, the upper bound
on the branching ratio of $\mu\to e\gamma$ is found to be stringent
\cite{meg}. Moreover, we have $Br(\mu\to e\bar{\nu}_e\nu_\mu)\approx 100\%$.
Using this and neglecting the electron mass, the branching ratio of
$\mu\to e\gamma$ is found to be
\begin{eqnarray}
Br(\mu\to e\gamma)&=&\frac{3\alpha}{16\pi G_F^2}
\left|\sum_{k=1}^3(Y_\nu)_{1k}(Y_\nu)_{2k}\times \right.
\nonumber \\
&& \left.
\left\{\frac{1}{4\mu_\eta^2}\left[f_2(x_{Rk})+f_2(x_{Ik})\right]
-\left[\cos^2\theta\frac{f_2(x_{k2})}{2m^2_{2+}}+\sin^2\theta\frac{f_2(x_{k1})}
{2m^2_{1+}}\right]\right\}\right|^2
\label{Eq:br}
\end{eqnarray}
Here, $\alpha =\frac{e^2}{4\pi}$ and $G_F$ is the Fermi constant. 

Here we compare our work with that of Ref.\cite{st}. The model in
\cite{st} is similar to that of \cite{ma}. But, in \cite{st}, a theory
at a high scale with an anomalous U(1)$_X$ symmetry is assumed.
The U(1)$_X$ symmetry breaks into $Z_2$ symmetry at a low scale.
Due to these differences,
there exists three 1-loop diagrams for neutrinos in \cite{st}, whereas
only two diagrams generate neutrino masses in \cite{ma}. The diagrams
for the LFV processes of $\ell_i\to\ell_j\gamma$ in \cite{st} is
similar to the diagrams given in this paper (see Figure 2).
But the expression for
$Br(\mu\to e\gamma)$, which is given in Eq. (\ref{Eq:br}), is found to
be different from that in \cite{st}. We hope that these differences might
have arised since the model in \cite{st} has different origin
from that of \cite{ma}.

Although the main motivation of this paper is to study the correlation
between neutrino masses and $Br(\mu\to e\gamma)$, below we mention about
muon $g-2$ in our model. It is known that the theoretical \cite{thg-2}
and experimental \cite{exg-2}
values of muon $g-2$ differ by about $3\sigma$ deviation. However,
there are hadronic uncertainities to muon $g-2$, which need to be
improved \cite{thg-2}. Hence, the above mentioned result is still an indication
for new physics signal. In our model \cite{ma}, muon $g-2$ get contributions
from MSSM fields \cite{mssmg-2} as well as from additional fields, which are
shown in Table 1. The contribution from MSSM fields can fit the $3\sigma$
discrepancy of muon $g-2$\footnote{In Ref.\cite{rsh}, the discrepancy
in muon $g-2$ is fitted in a supersymmetric model, where the contribution is
actually from the MSSM fields.}. Hence, in our model \cite{ma}, it is
interesting to know how large would be the contribution from the
additional fields of this model. The contribution from these additional fields
can be found from the amplitude of Eq. (\ref{Eq:amp}), which is given below.
\begin{equation}
\Delta a_\mu =\frac{m^2_\mu}{16\pi^2}\sum_{k=1}^3\left[(Y_\nu)_{2k}\right]^2
\left\{\frac{1}{2\mu_\eta^2}\left[f_2(x_{Rk})+f_2(x_{Ik})\right]
-\left[\cos^2\theta\frac{f_2(x_{k2})}{m^2_{2+}}+\sin^2\theta\frac{f_2(x_{k1})}
{m^2_{1+}}\right]\right\}
\label{Eq:damu}
\end{equation}
Here, $m_\mu$ is mass of the muon.

\section{Analysis and results}

As described in section 1 that our motivation is to study the correlation
between neutrino masses and $Br(\mu\to e\gamma)$. We have given expression
for neutrino masses in Eq. (\ref{Eq:mnu}). We have explained in the
previous section that to explain neutrino mass scale of 0.1 eV,
%by having SUSY particle masses around few 100 GeV, the
%neutrino Yukawa couplings, $(Y_\nu)_{ij}$, can be suppressed to
%${\cal O}(10^{-5})$. However, we have argued that alternatively
we can make neutrino Yukawa couplings to be about ${\cal O}(1)$, but we need to
fine-tune certain SUSY breaking soft parameters which are given in
Eq. (\ref{Eq:para}).
We consider this case, since for unsuppressed neutrino Yukawa couplings,
$Br(\mu\to e\gamma)$ can have maximum values. As mentioned before,
experiments have put the following upper bound: $Br(\mu\to e\gamma)<5.7\times
10^{-13}$ \cite{meg}.
Hence, for the above mentioned parameter space, where neutrino
Yukawa couplings are unsuppressed, we compute
$Br(\mu\to e\gamma)$ by fitting neutrino masses. We check if the computed
values for $Br(\mu\to e\gamma)$ satisfy the experimental bound \cite{meg}.

Before we compute $Br(\mu\to e\gamma)$,
we first need to ensure that the neutrino Yukawa couplings can be unsuppressed
in our model. We can calculate these Yukawa couplings from
Eq. (\ref{Eq:mnu}) by fitting to the neutrino oscillation data. The neutrino
mass matrix of Eq. (\ref{Eq:mnu}) is related to neutrino mass eigenvalues
through the following relation.
\begin{equation}
m_\nu=U_{\rm PMNS}^*{\rm diag}(m_1,m_2,m_3)U_{\rm PMNS}^\dagger.
\label{Eq:neumax}
\end{equation}
Here, $m_{1,2,3}$ are the mass eigenvalues of neutrinos and $U_{\rm PMNS}$
is the Pontecorvo-Maki-Nakagawa-Sakata matrix. The matrix $U_{\rm PMNS}$
depends on three mixing angles ($\theta_{12},\theta_{23},\theta_{13}$)
and Dirac CP-violating phase, $\delta_{\rm CP}$. In the above equation
there is a possibility of Majarona phases, which we have taken to be
zero, for simplicity. We have parametrized $U_{\rm PMNS}$ in terms of
mixing angles and $\delta_{\rm CP}$ as it is given in \cite{pdg}.

By fitting to various neutrino oscillation data, we haven known solar
and atmospheric neutrino mass-square differences and also about the
neutrino mixing angles \cite{neuos}. In the case of normal hierarchy (NH) of
neutrino masses, we have taken the mass-square differences as
\begin{equation}
\Delta m_{21}^2=m_2^2-m_1^2=7.6\times 10^{-5}~{\rm eV}^2,\quad
|\Delta m_{31}^2|=|m_3^2-m_1^2|=2.48\times 10^{-3}~{\rm eV}^2
\end{equation}
In the case of inverted hierarchy (IH) of neutrino masses, the value of
$\Delta m_{21}^2$ remains the same as mentioned above, but,
$|\Delta m_{31}^2|=2.38\times 10^{-3}~{\rm eV}^2$.
In this work, the neutrino mixing angles and CP-violating phase are
chosen to be
\begin{equation}
\sin\theta_{12}=\frac{1}{\sqrt{3}},\quad \sin\theta_{23}=\frac{1}{\sqrt{2}},
\quad \sin\theta_{13}=0.15,\quad \delta_{\rm CP}=0
\end{equation}
The above mentioned neutrino mass-square differences, mixing angles
and CP-violating phase are consistent with the fitted values in \cite{neuos}.
From the mass-square differences, we can estimate neutrino mass eigenvalues
which are given below for the cases of NH and IH, respectively.
\begin{eqnarray}
&& m_1=0,\quad m_2=\sqrt{\Delta m_{21}^2},\quad m_3=\sqrt{|\Delta m_{31}^2|}
\\
&& m_3=0,\quad m_1=\sqrt{|\Delta m_{31}^2|},\quad m_2=\sqrt{\Delta m_{21}^2
+m_1^2}
\end{eqnarray}

In the previous paragraph, we have mentioned neumerical values of
neutrino mass eigenvalues, mixing angles and CP-violating phase. By
plugging these values in Eq. (\ref{Eq:neumax}), we can compute the
elements of the matrix $m_\nu$, which are related to neutrino Yukawa
couplings and SUSY parameters through Eq. (\ref{Eq:mnu}).
Using Eq. (\ref{Eq:mnu}), we can calculate neutrino Yukawa couplings,
in order to satisfy neutrino oscillation data. This calculation procedure
would become simplified if we assume degenerate masses for right-handed
neutrinos and right-handed sneutrinos. For $i=1,2,3$, we assume the following:
\begin{equation}
M_i=M,\quad (m_N^2)_i=m_N^2,\quad (b_M)_i=b_M
\end{equation}
Under the above assumption, all the three right-handed neutrinos have mass
$M$. The corresponding sneutrinos have real and imaginary components
(see Eq. (\ref{Eq:sneu})), whose masses would be
\begin{equation}
m_R^2=M^2+m_N^2+b_M,\quad m_I^2=M^2+m_N^2-b_M
\end{equation}
Under the above mentioned assumption, the neutrino mass matrix of Eq.
(\ref{Eq:mnu}) will be simplified to
\begin{eqnarray}
(m_\nu)_{ij}&=&\frac{S_{ij}}{16\pi^2}\sum_{l=1}^3\left\{
M\left[[U_R(2,l)]^2\frac{m_{\eta_{Rl}}^2}{m_{\eta_{Rl}}^2-M^2}\ln
\frac{m_{\eta_{Rl}}^2}{M^2}-
[U_I(2,l)]^2\frac{m_{\eta_{Il}}^2}{m_{\eta_{Il}}^2-M^2}\ln
\frac{m_{\eta_{Il}}^2}{M^2}\right]+ \right.
\nonumber \\
&&\left. [U_\eta(2,l)]^2m_{\tilde{\eta}_l}
\left[\frac{m_{R}^2}{m_{R}^2-m_{\tilde{\eta}_l}^2}\ln\frac{m_{R}^2}
{m_{\tilde{\eta}_l}^2}-
\frac{m_{I}^2}{m_{I}^2-m_{\tilde{\eta}_l}^2}\ln\frac{m_{I}^2}
{m_{\tilde{\eta}_l}^2}\right]\right\},
\label{Eq:simmnu}
\\
S_{ij}&=&\sum_{k=1}^3(Y_\nu)_{ik}(Y_\nu)_{jk}
\label{Eq:S}
\end{eqnarray}
The elements $S_{ij}$ are expressed quadratic in neutrino Yukawa couplings.
From the above relation we can see that for certain values of SUSY parameters,
$S_{ij}$ can be calculated from $(m_\nu)_{ij}$.
Using the above mentioned assumption of
degenerate masses for right-handed neutrinos and right-handed
sneutrinos, we can see
that Eqs. (\ref{Eq:br}) $\&$ (\ref{Eq:damu}) would give us
$Br(\mu\to e\gamma)\propto S_{21}^2$ and $\Delta a_\mu\propto S_{22}$.
%We work in this assumption and study correlation between neutrino masses
%and $Br(\mu\to e\gamma)$, which we will address below.

In our model, there are plenty of SUSY parameters, and we need to fix
some of them to simplify our analysis.
In our analysis, we have chosen the following SUSY parameters as follows.
\begin{eqnarray}
&&\mu_\chi=600~{\rm GeV},\quad m_{\eta_1}=400~{\rm GeV},\quad
m_{\eta_2}=500~{\rm GeV},\quad m_\chi=600~{\rm GeV},
\nonumber \\
&&m_N=700~{\rm GeV},\quad \lambda_1=0.5,\quad \lambda_2=0.6,\quad \tan\beta=10
\end{eqnarray}
We have varied the parameters $\mu_\eta$ and $M$, freely. In the previous
section, we have explained that we need to fine-tune the parameters of
Eq. (\ref{Eq:para}) in order to get small neutrino masses. Among these
parameters, we take $(A\lambda)_1=\lambda_1\mu v_2/v_1$ and
$(A\lambda)_2=\lambda_2\mu v_1/v_2$. The other parameters of
Eq. (\ref{Eq:para}), without loss of generality, are taken to be degenerate,
which are given below.
\begin{equation}
b_M=b_\eta=b_\chi=b_{\rm susy}
\end{equation}

We have explained before that we have assumed degenerate masses for
right-handed neutrinos and right-handed sneutrinos.
Under this assumption, the information of neutrino Yukawa couplings is
contained in the quantities $S_{ij}$. Hence, it is worth to plot these
quantities to know about neutrino Yukawa couplings.
In Figure 3, for the case of NH,
we have plotted $S_{21}$ and $S_{22}$ versus
right-handed neutrino mass, $M$, for $\mu_\eta=$ 1 TeV.
\begin{figure}[!h]
\begin{center}

\includegraphics[height=2.5in,width=2.5in]{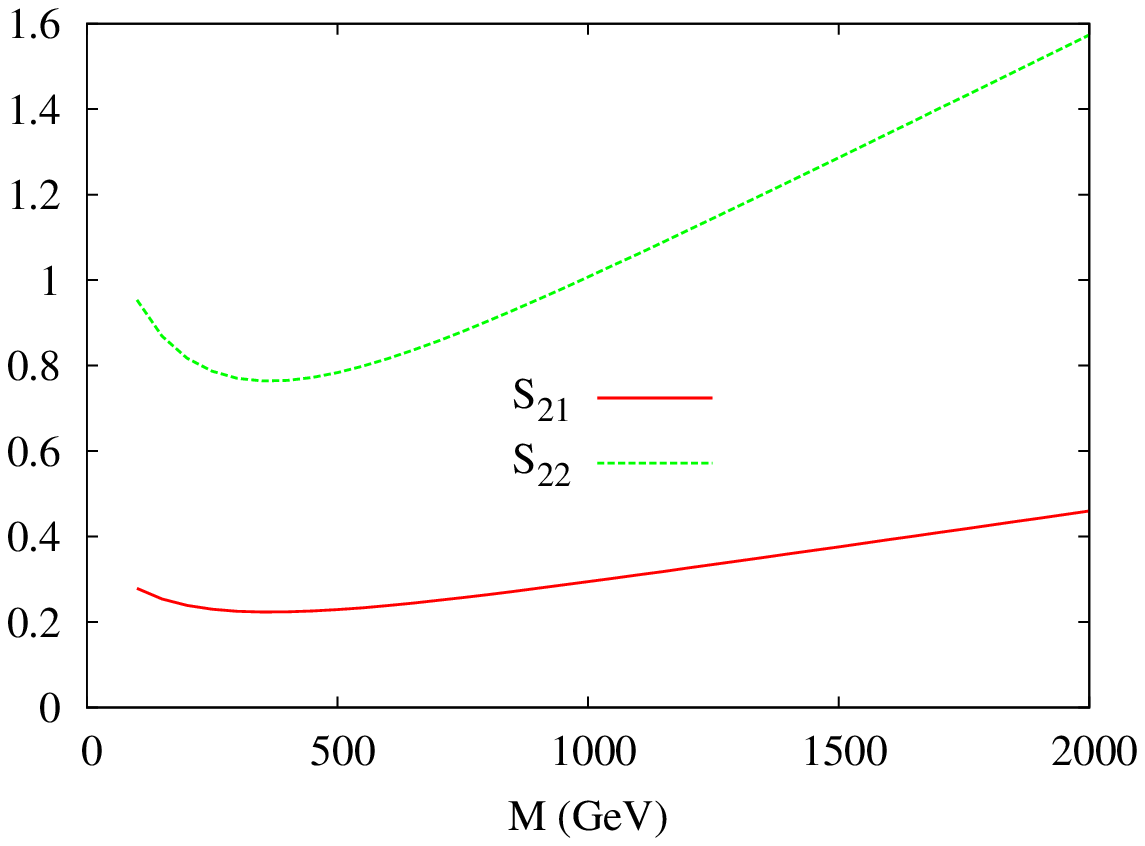}
\includegraphics[height=2.5in,width=2.5in]{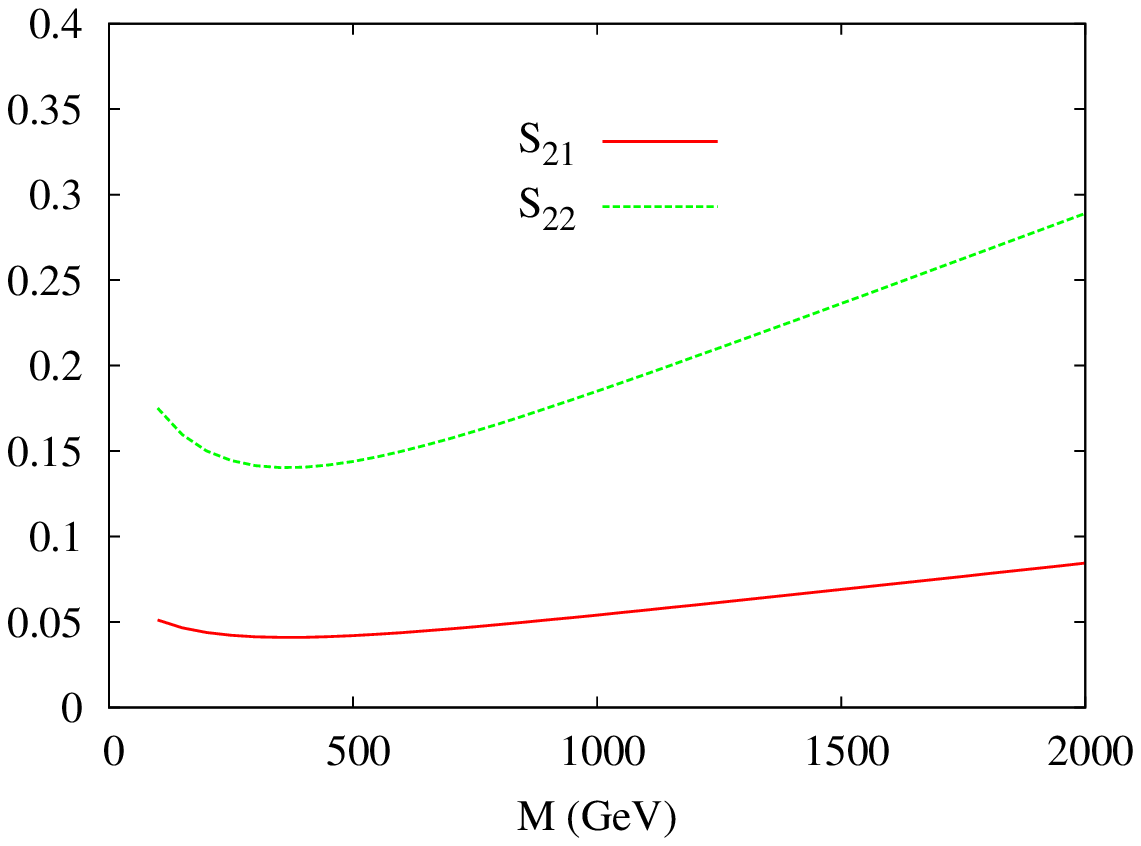}

\end{center}
\caption{The quantities $S_{21}$, $S_{22}$ are plotted against
right-handed neutrino mass for $\mu_\eta=$ 1 TeV, in the case of NH.
In the left-
and right-handed plots, $b_{\rm susy}$ is taken to be
$(3\times 10^{-2})^2$ GeV$^2$ and $(7\times 10^{-2})^2$ GeV$^2$, respectively.}
\end{figure}
The plots of Figure 3 indicate that $S_{22}$ and $S_{21}$ are around
${\cal O}(1)$. Since these quantities are sum of squares of neutrino Yukawa
couplings (see, Eq. (\ref{Eq:S})), we can expect that
the neutrino Yukawa couplings should be in the range of ${\cal O}(1)$.
We have not plotted the values of $S_{11}$, $S_{31}$, etc in Figure 3,
but we have found that these will also be around ${\cal O}(1)$. We have
plotted $S_{21}$ and $S_{22}$ in Figure 3, since these two determine
$Br(\mu\to e\gamma)$ and $\Delta a_\mu$.

From the plots of Figure 3, we can notice that the values of $S_{22}$ are
higher than that of $S_{21}$. This fact follows from Eq. (\ref{Eq:simmnu}),
where we can see that $S_{ij}$ are proportional to $(m_\nu)_{ij}$,
which are determined by neutrino oscillation parameters. In the case of
NH, we have seen that $(m_\nu)_{22}$ is greater than $(m_\nu)_{21}$ by a
factor of 3.4, hence $S_{22}$ is always found to be larger than $S_{21}$.
It is clear from the plots of Figure 3 that by increasing $b_{\rm susy}$,
$S_{21}$ and $S_{22}$ would decrease. Again, this feature can be understood
from Eq. (\ref{Eq:simmnu}). As explained in the previous section, the
square brackets of Eq. (\ref{Eq:simmnu}) would tend to zero in the limit
$b_{\rm susy}\to 0$. So for large value of $b_{\rm susy}$ there will be
less partial cancellation in the square brackets, and hence $S_{21}$ and
$S_{22}$ would decrease. In both the plots of Figure 3, it is found that
the values of
$S_{21}$ and $S_{22}$ initially decreases with $M$, goes to a minima and then
increases. The shape of these curves can be understood by applying the
approximation of $\frac{b_{\rm susy}}{M^2}\ll 1$ in Eq. (\ref{Eq:simmnu}).
In the limit $b_{\rm susy}\to 0$, we can take
\begin{equation}
m_{\eta_{Rl}}^2=m^2_{\eta_l}(1+\delta_{Rl}),\quad
m_{\eta_{Il}}^2=m^2_{\eta_l}(1+\delta_{Il}),\quad
U_R(2,l)\approx U_I(2,l)=U_0(2,l)
\end{equation}
Here, $\delta_{Rl},\delta_{Il}\ll 1$.
Using the above mentioned approximations in Eq. (\ref{Eq:simmnu}), we get
\begin{eqnarray}
(m_\nu)_{ij}&=&\frac{S_{ij}}{16\pi^2}\sum_{l=1}^3\left\{[U_0(2,l)]^2
(\delta_{Rl}-\delta_{Il})M\frac{m^2_{\eta_l}}{m^2_{\eta_l}-M^2}
\left[1-\frac{M^2}{m^2_{\eta_l}-M^2}\ln\frac{m^2_{\eta_l}}{M^2}\right]+\right.
\nonumber \\
&& \left. [U_\eta(2,l)]^2m_{\tilde{\eta}_l}\frac{2b_{\rm susy}}
{M^2+m_N^2-m^2_{\tilde{\eta}_l}}\left[1-\frac{m^2_{\tilde{\eta}_l}}
{M^2+m_N^2-m^2_{\tilde{\eta}_l}}\ln\frac{M^2+m_N^2}{m^2_{\tilde{\eta}_l}}
\right]\right\}
\end{eqnarray}
In the summation of the above equation, the first and second lines arise
due to left- and right-handed diagrams of Figure 1. From the above equation,
we can understand that the contribution from the first line increases,
reaches a maximum, and then decreases with $M$. Whereas, the contribution
from the second line of the above equation decreases monotonically with $M$.
It is this functional dependence on $M$ that determine the shape of the
lines in Figure 3.
Physically, in the limit $b_{\rm susy}\to 0$, the above description
suggests that the right-handed diagram of Figure 1 is significant only for
very low values of $M$. For other values of $M$, the left-handed diagram
of Figure 1 gives dominant contribution to neutrino masses. One
remark about the plots in Figure 3 is that we have fixed $\mu_\eta=$ 1 TeV
in these figures. We have varied $\mu_\eta$ from 500 GeV to 1.5 TeV and
have found that the plots in Figure 3 would change quantitatively, but
qualitative features would remain same. Also, the plots in Figure 3
are for the case of NH. Again, these plots can change quantitatively,
if not qualitatively, for the case of IH. For this reason, below we
present our results on $Br(\mu\to e\gamma)$ and muon $g-2$ for the
case of NH only.

As described before that our motivation is to compute $Br(\mu\to e\gamma)$
in the model of \cite{ma}. In Figure 3 we have shown that the neutrino
Yukawa couplings in this model can be ${\cal O}(1)$, and for these values
of Yukawa couplings, $Br(\mu\to e\gamma)$ is unsuppressed. In the parameter
space where neutrino Yukawa couplings are unsuppressed, we have
plotted $Br(\mu\to e\gamma)$ as a function of right-handed neutrino mass.
These plots are shown in Figure 4, where we have also varied $\mu_\eta$
from 500 GeV to 1.5 TeV.
\begin{figure}[!h]
\begin{center}

\includegraphics[height=2.5in,width=2.5in]{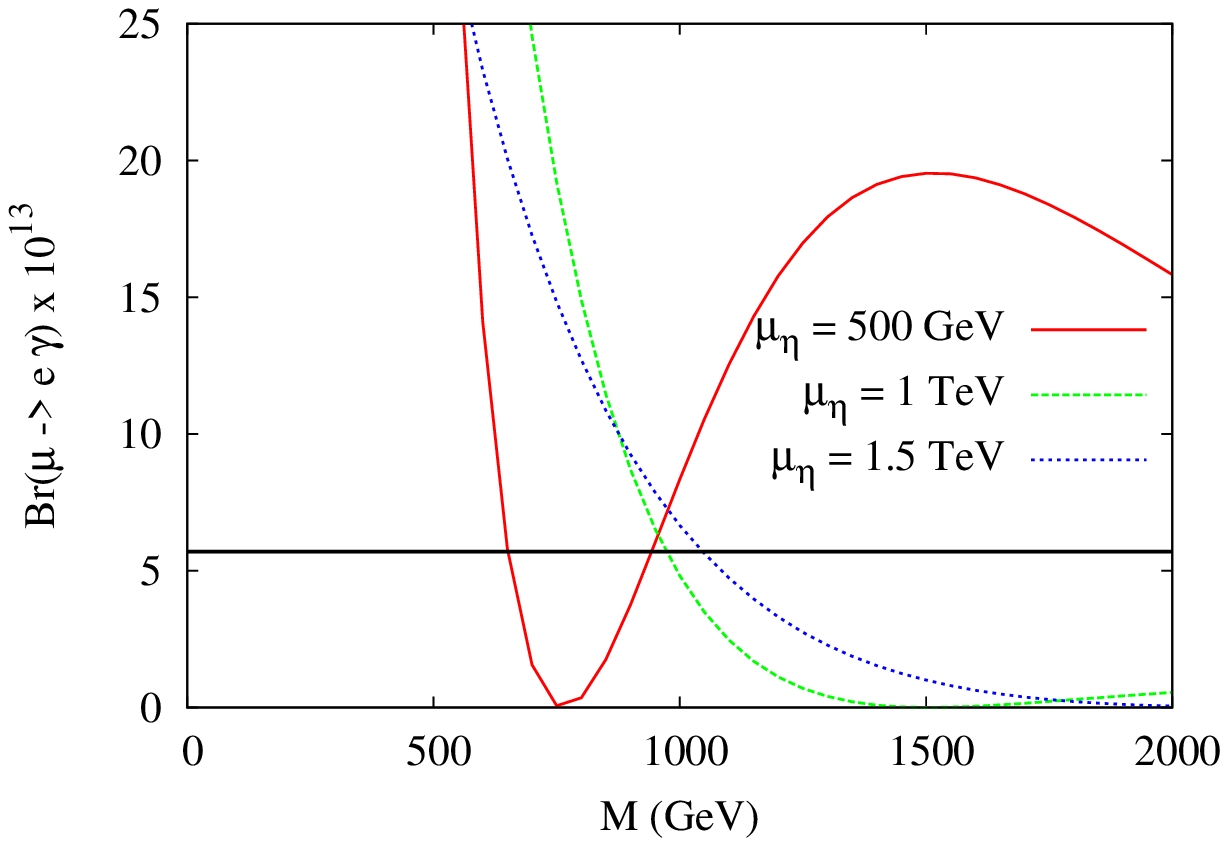}
\includegraphics[height=2.5in,width=2.5in]{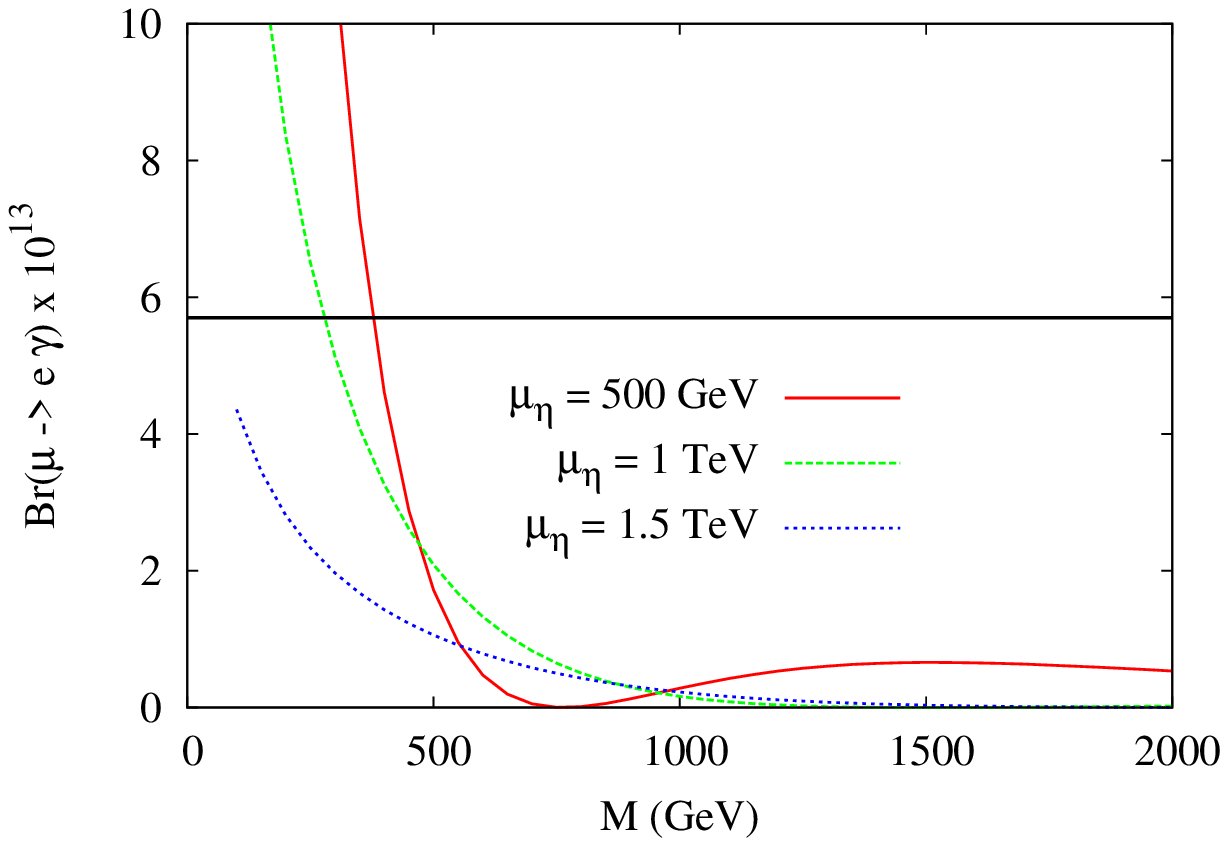}

\end{center}
\caption{$Br(\mu\to e\gamma)$ is plotted against
right-handed neutrino mass for different values of $\mu_\eta$. In the left-
and right-handed plots, $b_{\rm susy}$ has been taken as
$(3\times 10^{-2})^2$ GeV$^2$ and $(7\times 10^{-2})^2$ GeV$^2$, respectively.
The horizontal line indicates the current upper bound on
$Br(\mu\to e\gamma)$.}
\end{figure}
The horizontal line in these plots indicate the current upper bound of
$Br(\mu\to e\gamma)<5.7\times 10^{-13}$. This upper bound would impose
lower bound on the right-handed neutrino mass, as can be seen in the
plots of Figure 4. In the left-handed plot of Figure 4, for $\mu_\eta=$
500 GeV, the right-handed neutrino mass is allowed to be between
about 650 to 950 GeV. In the same plot, for $\mu_\eta=$ 1 or 1.5 TeV,
the right-handed neutrino mass has a lower bound of about 1 TeV.
In the right-handed plot of Figure 4, the lower bound on right-handed
neutrino mass is within 500 GeV, even for a low value of $\mu_\eta=$
500 GeV.

The lower bounds on the right-handed neutrino mass, $M$, are severe in
the left-handed plot of Figure 4. The reason is that for low value of
$b_{\rm susy}$, $S_{21}$ would be high, and hence $Br(\mu\to e\gamma)$ would
be large.
%From Figure 4, we can observe that lower bound on $M$ can be
%less severe by increasing $\mu_\eta$. We can understand this behaviour
%by noticing that the decay $\mu\to e\gamma$ is governed by propagating
%particles such as $\tilde{\eta}^+$ and $\eta_{m1,2}^+$, as can be
%seen from Figure 2. The masses of these propagating particles are proportional
%to $\mu_\eta$. Hence, $Br(\mu\to e\gamma)$ would be suppressed with
%increasing $\mu_\eta$, and as a result, we get less severe bounds on the
%right-handed neutrino mass.
From Figure 4, we can observe that $Br(\mu\to e\gamma)$ initially
decreases with $M$, goes to a minimum and then increases.
For instance, in the left-handed plot of Figure 4, for
$\mu_\eta=$ 500 GeV, $Br(\mu\to e\gamma)$ goes to a minimum around $M=$
750 GeV, and then it will have a local maxima around $M=$ 1.5 TeV. The
reason for $Br(\mu\to e\gamma)$ to initially decrease with $M$ is due
to the fact that the decay $\mu\to e\gamma$ is driven by right-handed
neutrinos and right-handed sneutrinos, as given in Figure 2.
The masses of right-handed
neutrinos and right-handed sneutrinos are proportional to $M$, and hence
$Br(\mu\to e\gamma)$ would be suppressed with increasing $M$. After that,
at a certain value of $M$, $Br(\mu\to e\gamma)$ would tend to become zero. The
reason for this is that the sum of the two diagrams of Figure 2 gives
a relative minus sign to the contribution of $Br(\mu\to e\gamma)$,
which is given in Eq. (\ref{Eq:br}). Hence, for a particular value of
$M$, the contributions from both the two diagrams of Figure 2 cancel out
and give a minimum for $Br(\mu\to e\gamma)$. Also, $Br(\mu\to e\gamma)$
can go to zero asymptotically when $M\to\infty$, since in this limit
the masses of right-handed neutrinos and right-handed sneutrinos would become
infinitely large and suppress $Br(\mu\to e\gamma)$. Hence,
$Br(\mu\to e\gamma)$ has two zeros on the $M$-axis. As $Br(\mu\to e\gamma)$
is a continous function of $M$ and is always a positive quantity,
it is having a local maxima between the two zeros on the $M$-axis.

In the previous section we have described about muon $g-2$. In Eq.
(\ref{Eq:damu}), we have given the contribution due to additional
fields (see Table 1) of our model to the muon $g-2$. Apart from this
contribution, MSSM fields of our model also contribute to muon $g-2$
\cite{mssmg-2}, and it is known that this contribution fits the
$3\sigma$ discrepancy of muon $g-2$. Hence, it is interesting to know
if the additional contribution of Eq. (\ref{Eq:damu}) could
be as large as that of MSSM contribution to muon $g-2$. In Figure 5,
we have plotted the contribution of Eq. (\ref{Eq:damu}). In the plots
of Figure 5, we have chosen the parameter region such that the neutrino
oscillation data is fitted.
\begin{figure}[!h]
\begin{center}

\includegraphics[height=2.5in,width=2.5in]{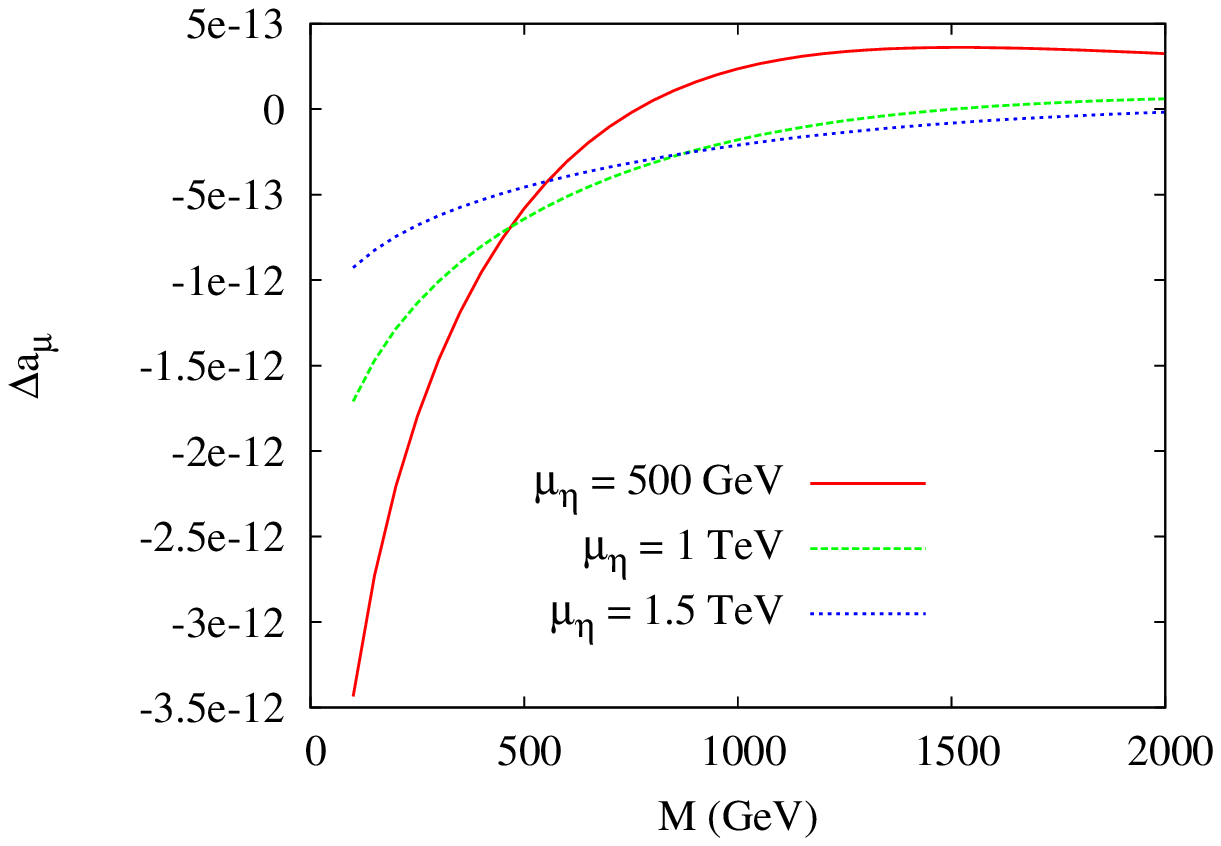}
\includegraphics[height=2.5in,width=2.5in]{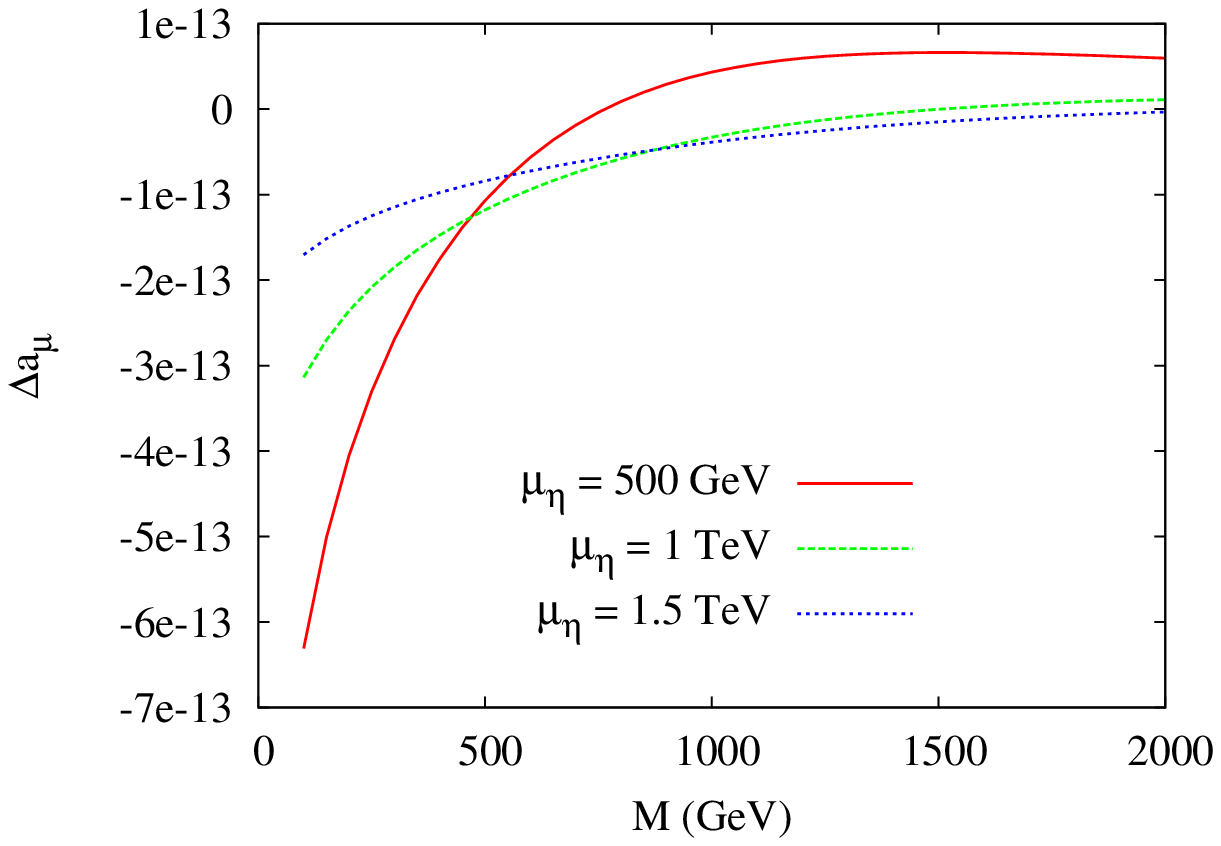}

\end{center}
\caption{$\Delta a_\mu$ is plotted against
right-handed neutrino mass for different values of $\mu_\eta$. In the left-
and right-handed plots, $b_{\rm susy}$ has been taken as
$(3\times 10^{-2})^2$ GeV$^2$ and $(7\times 10^{-2})^2$ GeV$^2$, respectively.}
\end{figure}
From the plots of Figure 5, we can see that for low values of $M$,
$\Delta a_\mu$ can be negative and it becomes positive after certain large
value of $M$. From these plots we can notice that the overall magnitude of
$\Delta a_\mu$ is not more than about $10^{-12}$. This contribution is atleast
two orders smaller than the estimated discrepancy of muon $g-2$, which is
$(29\pm 9)\times 10^{-10}$ \cite{thg-2}. From this we can conclude that
the additional contribution to muon $g-2$ in our model, {\it i.e.}
Eq. (\ref{Eq:damu}), is insignificant compared to the MSSM contribution
to muon $g-2$.

\section{Conclusions}

We have worked in a supersymmetric model where neutrino masses arise at
1-loop level \cite{ma}. We have computed these loop diagrams and obtained
expressions for neutrino masses. We have identified a parameter region of
this model, where the neutrino osicllation data can be fitted without
the need of suppressing the neutrino Yukawa couplings. In our parameter
region, the SUSY breaking soft parameters such as $b_M$, $b_\eta$,
$b_\chi$, $(A\lambda)_1$,
$(A\lambda)_2$ need to be fine-tuned. In this parameter region, branching
fraction of $\mu\to e\gamma$ can be unsuppressed, and hence, we have computed
$Br(\mu\to e\gamma)$. We have shown that the current upper bound on
$Br(\mu\to e\gamma)$ can put lower bounds on the mass of right-handed
neutrino field. Depending on the parameteric choice, we have found that this
lower bound can be about 1 TeV. We have also computed the contribution to muon
$g-2$ arising from additional fields of this model, which are given in
Table 1. We have shown that, in the region where neutrino oscillation
data is fitted, the above mentioned contribution is two orders smaller than
the discrepancy in muon $g-2$.


\begin{thebibliography}{99}

\bibitem{bsm}
C. Quigg, arXiv:hep-ph/0404228;\\
J. Ellis, Nucl. Phys. A {\bf 827} (2009) 187C [arXiv:0902.0357 [hep-ph]].

\bibitem{neurev}
%For a review on neutrino masses and mixing, see \\
R. N. Mohapatra, arXiv:hep-ph/0211252;\\
Y. Grossman, arXiv:hep-ph/0305245;\\
A. Strumia and F. Vissani, arXiv:hep-ph/0606054.

\bibitem{susy}
  H.~P.~Nilles, Phys. Rept. {\bf 110}, (1984) 1;\\
  H.~E.~Haber and G.~L.~Kane, Phys. Rept. {\bf 117}, (1985) 75;\\
  S.~P.~Martin, arXiv:hep-ph/9709356;\\
  M.~Drees, R.~Godbole and P.~Roy, Theory and Phenomenology of Sparticles,
  (World Scientific, 2004);\\
  P.~Binetruy, Supersymmetry (Oxford University Press, 2006);\\
  H.~Baer and X.~Tata, Weak Scale Supersymmetry: From
  Superfields to Scattering Events, (Cambridge University Press, 2006).

\bibitem{lfvrev}
T. Mori, eConf C {\bf 060409}, (2006) 034 [hep-ex/0605116];\\
J. M. Yang, Int. J. Mod. Phys. A {\bf 23}, (2008) 3343 (2008) [arXiv:0801.0210 [hep-ph]];\\
A. J. Buras, Acta Phys. Polon. Supp. {\bf 3}, (2010) 7 (2010) [arXiv:0910.1481 [hep-ph]];\\
Y. Nir, CERN Yellow Report CERN-2010-001, 279-314 [arXiv:1010.2666 [hep-ph]].

\bibitem{meg}
J. Adam {\it et al.} (MEG Collaboration), Phys. Rev. Lett. {\bf 110}
(2013) 201801.

\bibitem{lfvboun}
B. Aubert {\it et al.} (BaBar Collaboration), Phys. Rev. Lett. {\bf 104}
(2010) 021802.

\bibitem{pdg}
K.A. Olive {\it et al.} (Particle Data Group), Chin. Phys. C {\bf 38} (2014)
090001.

\bibitem{ma}
E. Ma, Annales Fond. Broglie {\bf 31} (2006) 285, [arXiv:hep-ph/0607142].

\bibitem{scoto}
E. Ma, Phys. Rev. D {\bf 73} (2006) 077301.

\bibitem{IH}
R. Barbieri, L.J. Hall and V.S. Rychkov, Phys. Rev. D{\bf 74} (2006) 015007.

\bibitem{IHrec}
A. Arhrib, R. Benbrik, J.E. Falaki and A. Jueid, [arXiv:1507.03630];\\
A.D. Plascencia, JHEP 1509 (2015) 026;\\
S. Kashiwase and D. Suematsu, Phys. Lett. B{\bf 749} (2015) 603.

\bibitem{multi}
Q.-H. Cao, E. Ma, Jose Wudka and C.-P. Yuan, arXiv:0711.3881.

\bibitem{refs}
H. Fukuoka, J. Kubo and D. Suematsu, Phys. Lett. B {\bf 678} (2009) 401;\\
D. Suematsu, T. Toma and T. Yoshida, Int. J. Mod. Phys. A{\bf 25} (2010) 4033.

\bibitem{st}
D. Suematsu and T. Toma, Nucl. Phys. B{\bf 847} (2011) 567.

\bibitem{su5}
E. Ma, Phys. Lett. B{\bf 659} (2008) 885.

\bibitem{u1ma}
E. Ma, Mod. Phys. Lett. A{\bf 23} (2008) 721.

\bibitem{ihl}
I-H. Lee, Phys. Lett. B{\bf 138} (1984) 121;
Nucl. Phys. B{\bf 246} (1984) 120.

\bibitem{mu3e}
U. Bellgardt {\it et al.} (SINDRUM Collaboration), Nucl. Phys. B{\bf 299}
(1988) 1.

\bibitem{akot}
M. Aoki, J. Kubo, T. Okawa and H. Takano, Phys. Lett. B{\bf 707} (2012) 107.

\bibitem{fulfv}
A.M. Baldini {\it et al.}, arXiv:1301.7225;\\
T. Aushev {\it et al.}, arXiv:1002.5012.

\bibitem{af}
A.J.R. Figueiredo, Eur. Phys. J. C{\bf 75} (2015) 3, 99.

\bibitem{thg-2}
F. Jegerlehner and A. Nyffeler, Phys. Rept. {\bf 477} (2009) 1;\\
T. Blum, A. Denig, I. Logashenko, E. de Rafael, B. Lee Roberts, T. Teubner
and G. Venanzoni, arXiv:1311.2198.

\bibitem{exg-2}
G.W. Bennett {\it et al.} (Muon g-2 Collaboration), Phys. Rev. D{\bf 73}
(2006) 072003.

\bibitem{mssmg-2}
T. Moroi, Phys. Rev. D{\bf 53} (1996) 6565 [Erratum-ibid. D{\bf 56}
(1997) 4424];\\
S.P. Martin and J.D. Wells, Phys. Rev. D{\bf 64} (2001) 035003.

\bibitem{rsh}
R.S. Hundi, Phys. Rev. D{\bf 83} (2011) 115019.

\bibitem{neuos}
D.V. Forero, M. Tortola and J.W.F. Valle, Phys. Rev. D{\bf 90} (2014) 093006.\\

\end{thebibliography}
\end{document}